\documentclass[aip,jcp,preprint]{revtex4-2}

\usepackage{amsmath,amssymb,mathtools,bm}
\usepackage{graphicx}
\usepackage{booktabs}
\usepackage{hyperref}
\hypersetup{hidelinks,
pdftitle={From Transition-State Geometry to Gap Times: What Lagrangian Betweenness Measures in Chemical Reaction Dynamics},
pdfauthor={Stephen Wiggins}}

\newcommand{\LB}{B^{\mathrm L}}
\newcommand{\dd}{\mathrm d}
\newcommand{\norm}[1]{\left\lVert #1\right\rVert}
\newcommand{\abs}[1]{\left\lvert #1\right\rvert}
\newcommand{\OO}{\mathcal{O}}

\begin{document}

\title{From Transition-State Geometry to Gap Times:\\
What Lagrangian Betweenness Measures in Chemical Reaction Dynamics}

\author{Stephen Wiggins}
\affiliation{Hetao Institute for Mathematics and Interdisciplinary Sciences (HIMIS), Shenzhen, China}
\affiliation{School of Mathematics, University of Bristol, Bristol BS8 1UG, United Kingdom}

\date{September 22, 2026}

\begin{abstract}
A chemical transition state is a phase-space bottleneck: trajectories approach along stable directions and leave along unstable directions. Lagrangian betweenness (LB), introduced in finite-time transport theory for fluid flows, combines backward and forward deformation in a way suggestive of this gather-and-disperse geometry. We ask what LB measures when the transition-state geometry is known. For a linear rank-one saddle, LB is spatially constant even though the stable and unstable manifolds are present. Relative nonlinear corrections remain small on shrinking initial neighborhoods, including observation times on the logarithmic local escape scale. At fixed spatial resolution, however, increasing observation time can concentrate the spatial variation of LB near a stable or unstable manifold. A separable quartic Hamiltonian makes these two limits explicit. A nonseparable Hamiltonian shows local flattening near a hyperbolic periodic orbit. We then use HCN/CNH isomerization, where dividing surfaces and gap times are established by phase-space transition-state theory. Before any exit, larger LB generally accompanies longer eventual gap times. The incoming reactive action provides a dynamical interpretation: initial conditions closer to the normally hyperbolic invariant manifold have longer local passages and greater accumulated stretching during passage. Among trajectories still inside at $0.5$ ps that exit before $5$ ps, early LB does not usefully rank exit times. LB therefore characterizes transition-state-organized transport in this example; a large value is neither an intrinsic manifold marker nor a general measure of molecular residence time.
\end{abstract}


\maketitle

\section{Introduction}
\label{sec:intro}

A chemical reaction is often geometrically depicted as motion over a barrier in configuration space. The dynamical picture is richer. Near a rank-one saddle, trajectories approach a small phase-space region along stable directions and leave it along unstable directions. A normally hyperbolic invariant manifold (NHIM) forms the core of this transition-state geometry, and a dividing surface anchored on the NHIM separates the two directions of passage. This phase-space formulation grew from the transition-state ideas of Eyring, Evans and Polanyi, and Wigner and from the subsequent dynamical development of transition-state theory \cite{Eyring1935,EvansPolanyi1935,Wigner1938,PollakPechukas1978,Pechukas1981,WigginsWiesenfeldJaffeUzer2001,UzerJaffePalacianYanguasWiggins2002,WaalkensBurbanksWiggins2004,WaalkensSchubertWiggins2008,MacKayStrub2014,Wiggins1994,WigginsReactionBook2025}.

That geometry gives us a useful standard against which to judge finite-time diagnostics. If the NHIM, its stable and unstable manifolds, and the dividing surface are already known from phase-space transition-state theory, then the value of a new scalar diagnostic need not lie in reconstructing that geometry. The more informative question is: \emph{what feature of the trajectories organized by that geometry does the diagnostic detect?}

Finite-time Lyapunov exponents (FTLEs) measure the growth of infinitesimal perturbations over a prescribed time interval and have played an important role in finite-time transport theory \cite{ShaddenLekienMarsden2005,Haller2015}. Their interpretation requires care: a hyperbolic invariant set need not produce a spatially varying FTLE field, and a region in which the FTLE is large need not by itself identify an invariant manifold \cite{BranickiWiggins2010}. This point is especially relevant here because the LB construction used by García-Cuadrillero \emph{et al.} is built directly from forward and backward finite-time stretching rates.

Lagrangian descriptors (LDs) use trajectory integrals to reveal finite-time phase-space structure \cite{JimenezMadridMancho2009,MendozaMancho2010,ManchoWigginsCurbeloMendoza2013}. For a widely used class of LDs, the important saddle signature is not simply a large value but a loss of differentiability across the stable and unstable manifolds in suitably aligned coordinates. This mechanism has been analyzed mathematically and tested in reaction-dynamics examples where the invariant manifolds are already known \cite{LopesinoEtAl2017,NaikGarciaGarridoWiggins2019}. Revuelta, Benito, and Borondo used this framework for LiCN/LiNC isomerization and emphasized the role of integration time in making the transition-state structure visible \cite{RevueltaBenitoBorondo2021}.

Lagrangian betweenness (LB) comes from a different line of work. Betweenness centrality in network theory measures how strongly a node lies between many source--destination paths \cite{Freeman1977}. Flow-network methods translated related ideas into transport between spatial regions in fluid flows \cite{SerGiacomiEtAl2015}, and Ser-Giacomi and co-workers introduced LB as a continuous finite-time analogue that combines information from the past and the future \cite{SerGiacomiEtAl2021}. García-Cuadrillero, Revuelta, Benito, and Borondo recently brought LB into chemical reaction dynamics in a detailed study of LiCN/LiNC isomerization \cite{GarciaCuadrilleroEtAl2026}. At short times they found little transition-state structure, at intermediate times a clear cross associated with the transition state and its invariant manifolds, and at longer times additional global structures.

The present paper is motivated by that observation. We ask what dynamical mechanism causes the LB field to develop spatial structure associated with a transition-state manifold, and what dynamical information that structure contains. To avoid circular validation, we use systems for which the relevant transition-state geometry is already known from independent dynamical constructions.

Two limits will be important. If the observation time is fixed while the spatial window shrinks toward the saddle equilibrium, any smooth finite-time scalar field becomes locally flat. The stronger result established below is that the relative spatial variation of LB still vanishes when the observation time grows on the logarithmic local escape scale, provided the forward and backward histories remain in the region controlled by the local estimate. The complementary limit is equally important: if the spatial resolution is held fixed while the observation time increases, the spatial variation of LB can become increasingly concentrated in a narrow neighborhood of a stable or unstable manifold. These two statements are compatible and explain why increasing the integration time can improve the visibility of transition-state structure at finite resolution without implying that LB loses differentiability on the manifold.

The second part of the paper asks a chemical question. For an initial condition chosen on the incoming dividing surface, the \emph{gap time} is the time required for its trajectory to reach an outgoing dividing surface. Gap-time distributions retain short passages, returns, multiple dynamical populations, and long-lived tails and have a long history in unimolecular reaction theory \cite{Thiele1962,BrumerFitzWardlaw1980,DeLeonBerne1981,DumontBrumer1986,WaalkensBurbanksWiggins2005,EzraWaalkensWiggins2009}. We use the HCN/CNH isomerization model because its phase-space transition-state geometry, dividing surfaces, normal-form maps, and gap-time statistics are already well established. This lets us ask what aspects of molecular residence LB detects without using an LB calculation to validate itself. The normal-form construction, trajectory calculations, and numerical tests reported here are documented in a public reproducibility package \cite{WigginsRepro2026}.

The organization follows that logic. Section~\ref{sec:definitions} reviews the existing LB construction and gives its finite-time deformation interpretation. Section~\ref{sec:localtheory} calculates LB near a rank-one saddle and then introduces LDs to expose the different local mechanisms of the two diagnostics. Section~\ref{sec:benchmarks} uses two model Hamiltonians to separate shrinking spatial scale and increasing observation time; the coupled calculation in Sec.~\ref{sec:floquetcheck} then tests whether the local flattening persists when separability is lost. Sections~\ref{sec:hcnmodel} and \ref{sec:hcnlocal} introduce HCN/CNH and test the saddle-equilibrium prediction in the molecular Hamiltonian. Section~\ref{sec:gaptimes} then asks why pre-exit LB is strongly associated with the later gap time and separates transition-state-organized passage from subsequent trapping. Section~\ref{sec:discussion} discusses the interpretation and conclusions.

\section{What Lagrangian betweenness measures}
\label{sec:definitions}

This section reviews the Lagrangian-betweenness construction introduced by Ser-Giacomi \emph{et al.} and subsequently used by García-Cuadrillero \emph{et al.} for LiCN/LiNC isomerization \cite{SerGiacomiEtAl2021,GarciaCuadrilleroEtAl2026}. No new definition of LB is introduced here. Our purpose is to state the existing construction in a form that makes its dynamical meaning transparent before applying it to chemical transition-state geometry.

\subsection{From trajectories to finite-time deformation}

Consider a time-dependent differential equation
\begin{equation}
 \dot z=f(z,t),\qquad z\in\mathbb{R}^m.
 \label{eq:nonautonomous}
\end{equation}
The vector $z$ specifies a point in phase space. If the system is at $z_0$ at time $t_0$, denote its state at time $t$ by
\begin{equation}
 \Phi(t,t_0;z_0)=z(t;t_0,z_0).
 \label{eq:process}
\end{equation}
The map $\Phi$ is the finite-time evolution of the state. For an autonomous system it depends only on the elapsed time, and we write $\Phi^T(z_0)$ with $T=t-t_0$. We retain the two-time notation briefly because LB was introduced for finite-time transport in time-dependent flows \cite{KloedenRasmussen2011,SerGiacomiEtAl2021}.

The trajectory tells us where a point goes. LB instead uses how nearby points separate. The derivative of the finite-time map with respect to its initial condition is the tangent map
\begin{equation}
 M(t,t_0;z_0)=D_{z_0}\Phi(t,t_0;z_0).
 \label{eq:tangentmap}
\end{equation}
Choose a fixed inner product in phase space and its associated norm; in the coordinates used below this is the Euclidean norm unless stated otherwise. For any matrix $M$, define
\begin{equation}
 \sigma_{\max}(M)=\max_{\norm v=1}\norm{Mv}.
 \label{eq:sigmamax}
\end{equation}
Thus $\sigma_{\max}(M)$ is the largest factor by which the tangent map can stretch an infinitesimal unit displacement.

For a duration $T\ge0$, define the forward and backward stretching factors
\begin{equation}
\begin{aligned}
 S_+(z,t;T)&=\sigma_{\max}\!\left(D_z\Phi(t+T,t;z)\right),\\
 S_-(z,t;T)&=\sigma_{\max}\!\left(D_z\Phi(t-T,t;z)\right).
\end{aligned}
 \label{eq:stretchfactors}
\end{equation}
The first follows a small displacement into the future; the second follows it into the past. For $T>0$, the corresponding maximum-stretching finite-time Lyapunov exponents are
\begin{equation}
 \lambda_f(z,t;T)=\frac{1}{T}\ln S_+(z,t;T),\qquad
 \lambda_b(z,t;T)=\frac{1}{T}\ln S_-(z,t;T).
 \label{eq:ftledefs}
\end{equation}
Thus $S_+=e^{T\lambda_f}$ and $S_-=e^{T\lambda_b}$. FTLEs and their relation to finite-time transport have a substantial literature \cite{ShaddenLekienMarsden2005,Haller2015,BranickiWiggins2010}; here we introduce them only because they are the tangent-stretching ingredients from which LB is constructed.

\subsection{From network betweenness to a continuous-flow quantity}

Ordinary network betweenness begins with a simple idea. A node has high betweenness when many paths connecting different origins and destinations pass through it \cite{Freeman1977}. In a continuous flow there is no finite list of graph paths to count. Ser-Giacomi and co-workers therefore replaced the path count by finite-time deformation: large backward stretching serves as a local proxy for a diverse finite-time past, and large forward stretching as a proxy for a diverse finite-time future \cite{SerGiacomiEtAl2015,SerGiacomiEtAl2021}.

For Hamiltonian dynamics this interpretation requires one qualification. Liouville's theorem preserves phase-space volume. A small phase-space region therefore does not decrease in volume as it approaches a bottleneck. Its shape, however, can change substantially: a compact region can be stretched into a long, thin region. At a chosen finite resolution, the largest singular value measures the greatest linear stretching of that region. In a rank-one hyperbolic region, where one stretching direction dominates, this provides a local measure of how broadly its past or future is dispersed. It is not a literal count of trajectories and does not represent compression of phase-space volume.

This also explains why the LB construction is particularly natural near a rank-one bottleneck \cite{SerGiacomiEtAl2021}. If several independent directions expand comparably, a quantity based only on the largest singular value records only the dominant one.

\subsection{Why the backward and forward factors are multiplied and averaged}

Fix a total observation time $\tau>0$. Let $s$ denote a possible division of that interval into a past part of duration $s$ and a future part of duration $\tau-s$. For that split, define
\begin{equation}
 \beta(z;t_0,s,\tau)
 =S_-(z,t_0+s;s)\,S_+(z,t_0+s;\tau-s).
 \label{eq:splitproduct}
\end{equation}
The product combines backward and forward stretching and increases when either factor increases. It therefore rewards simultaneous amplification in both time directions, but a large product alone does not establish that both factors are large.

No split time is preferred, so the product is averaged over $0\le s\le\tau$:
\begin{equation}
 \LB(z;t_0,\tau)
 =\frac{1}{\tau}\int_0^\tau \beta(z;t_0,s,\tau)\,\dd s.
 \label{eq:LBprocess}
\end{equation}
For an autonomous system this becomes
\begin{equation}
 \LB(z;\tau)=\frac{1}{\tau}\int_0^\tau
 S_-(z;s)\,S_+(z;\tau-s)\,\dd s.
 \label{eq:LB}
\end{equation}
Using Eq.~\eqref{eq:ftledefs}, the split product may equivalently be written
\begin{equation}
 \beta
 =\exp\!\left[s\,\lambda_b(z,t_0+s;s)
 +(\tau-s)\lambda_f(z,t_0+s;\tau-s)\right].
 \label{eq:LBftleform}
\end{equation}
This is the form used by García-Cuadrillero \emph{et al.} \cite{GarciaCuadrilleroEtAl2026}. LB therefore contains no tangent information beyond the family of forward and backward finite-time stretching rates entering the split-time integral. Its distinctive content is the nonlinear way in which those factors are multiplied and averaged over all divisions of the observation interval. Appendix~\ref{app:numerics} checks whether that combination behaves differently from the one-sided stretching measures in the HCN/CNH ensemble.

Near a rank-one chemical bottleneck, stable motion supplies the gathering side and unstable motion the dispersal side. Whether this geometry produces a maximum or a loss of differentiability in LB must be determined from the dynamics.

\subsection{The norm is part of the definition}
\label{sec:metric}

The singular values in Eq.~\eqref{eq:stretchfactors} depend on the norm used to measure tangent vectors. A canonical change of coordinates preserves the symplectic structure but need not preserve Euclidean lengths. If the Euclidean norm is imposed independently before and after such a change, the numerical singular values and therefore the numerical LB values generally change \cite{LekienRoss2010}. We state the norm used in the molecular calculations and later repeat the principal HCN/CNH comparison in a second natural coordinate system.

Smoothness also requires care. A simple largest singular value, meaning one separated from the others, varies smoothly with a smooth tangent map. When competing singular values cross, that conclusion need not hold. The smoothness statements below are local, in regions without such crossings for positive integration durations. The local bound in Sec.~\ref{sec:localtheory} controls changes in the singular value directly and does not require differentiating it.

\section{Lagrangian betweenness near a rank-one saddle}
\label{sec:localtheory}

Near an index-one saddle of the potential-energy surface, the Hamiltonian has one hyperbolic degree of freedom and the remaining directions are elliptic. On an energy surface above the saddle energy, this local geometry gives rise to a normally hyperbolic invariant manifold and to a dividing surface anchored on that NHIM. We first ask what the LB construction gives in the quadratic approximation to this local saddle dynamics.

\subsection{The linear saddle--center flow}

For two degrees of freedom, write the quadratic Hamiltonian as
\begin{equation}
 H_2=E_s+\lambda qp+\frac{\omega}{2}(x^2+y^2),
 \qquad \lambda,\omega>0.
 \label{eq:H2}
\end{equation}
The pair $(q,p)$ forms the two-dimensional \emph{hyperbolic plane}: one direction expands exponentially and the other contracts exponentially. The pair $(x,y)$ forms the \emph{center plane}: the motion there is oscillatory. The saddle equilibrium is at $q=p=x=y=0$ and has energy $E_s$. On an energy surface $E>E_s$, the condition $q=p=0$ defines the local NHIM; for this two-degree-of-freedom model it is a periodic orbit in the center plane.

Hamilton's equations are
\begin{equation}
 \dot q=\lambda q,\qquad
 \dot p=-\lambda p,\qquad
 \dot x=\omega y,\qquad
 \dot y=-\omega x.
 \label{eq:linearfloweqs}
\end{equation}
Thus a trajectory beginning at $(q_0,p_0,x_0,y_0)$ satisfies
\begin{equation}
 q(T)=e^{\lambda T}q_0,\qquad
 p(T)=e^{-\lambda T}p_0,
 \label{eq:hyperflow}
\end{equation}
while
\begin{equation}
 \begin{pmatrix}x(T)\\y(T)\end{pmatrix}
 =R_{\omega T}\begin{pmatrix}x_0\\y_0\end{pmatrix},
 \qquad
 R_{\omega T}=\begin{pmatrix}
 \cos\omega T&\sin\omega T\\
 -\sin\omega T&\cos\omega T
 \end{pmatrix}.
 \label{eq:centerrotation}
\end{equation}
Thus $R_{\omega T}$ is simply the linear flow in the center plane: a rotation through angle $\omega T$.

Because the quadratic Hamiltonian generates a linear system, the derivative of the flow with respect to its initial condition is the same matrix that propagates the initial condition. Hence
\begin{equation}
 D\Phi^T
 =\operatorname{diag}\!\left(e^{\lambda T},e^{-\lambda T},R_{\omega T}\right).
 \label{eq:linearM}
\end{equation}
The rotation has singular values one. The largest singular value of the full forward map is therefore $e^{\lambda T}$, and the largest singular value of the backward map is also $e^{\lambda T}$. With more than two degrees of freedom, each additional center plane contributes another rotation block $R_{\omega_jT}$ with singular values one. The conclusion is therefore unchanged for one hyperbolic degree of freedom and any number of uncoupled linear center planes; this is the linear structure relevant to the three-degree-of-freedom HCN/CNH model used below.

In these coordinates, an initial condition with $q=0$ approaches the NHIM in forward time, so $q=0$ is the local stable manifold. An initial condition with $p=0$ approaches it in backward time, so $p=0$ is the local unstable manifold.

\subsection{LB is spatially constant for the linear saddle}

Consider a split $s$ of a total observation time $\tau$. Equation~\eqref{eq:linearM} gives
\begin{equation}
 S_-(z;s)S_+(z;\tau-s)
 =e^{\lambda s}e^{\lambda(\tau-s)}
 =e^{\lambda\tau}.
 \label{eq:linearsplit}
\end{equation}
The product is independent of the split time. More importantly, it is independent of the phase-space point. Averaging over $s$ gives
\begin{equation}
 \boxed{\LB(z;\tau)=e^{\lambda\tau}}.
 \label{eq:LBlinear}
\end{equation}
Ser-Giacomi \emph{et al.} already obtained the exponential expression for equal, constant forward and backward stretching rates [Eq.~(19) of Ref.~\cite{SerGiacomiEtAl2021}]. Here the linear saddle realizes those assumptions exactly, allowing the spatial interpretation to be examined against its known invariant manifolds.

The important feature of Eq.~\eqref{eq:LBlinear} is what does \emph{not} appear on its right-hand side: there is no dependence on $q$, $p$, $x$, or $y$. Every initial condition in the linear system receives exactly the same LB value for a given observation time. The stable manifold $q=0$, the unstable manifold $p=0$, and the NHIM $q=p=0$ are all present in the dynamics, but none is distinguished by spatial variation of LB because there is no spatial variation. Hyperbolicity by itself therefore does not force an LB feature on an invariant manifold.

\subsection{Why an LD behaves differently}

At this point a Lagrangian descriptor provides a useful comparison because its local manifold signature arises in a fundamentally different way. For a trajectory with components $z_j(t)$, consider the componentwise family
\begin{equation}
 M_\alpha(z_0;\tau)=\int_{-\tau}^{\tau}
 \sum_{j=1}^{m}\abs{\dot z_j(t;z_0)}^\alpha\,\dd t,
 \qquad 0<\alpha<1.
 \label{eq:LDdef}
\end{equation}
For the quadratic Hamiltonian, this separates into hyperbolic and center contributions,
\begin{equation}
 M_\alpha=M_\alpha^{\rm hyp}+M_\alpha^{\rm cen}.
 \label{eq:LDsplit}
\end{equation}
When we examine a crossing of the stable manifold $q=0$, we vary $q$ while holding the remaining initial coordinates fixed. The center contribution is then independent of the coordinate transverse to that manifold and cannot affect the corresponding transverse derivative. The same reasoning applies to a crossing of the unstable manifold $p=0$. We therefore examine the hyperbolic contribution for this particular question. This does not assert that the center contribution is globally smooth; it says only that it does not enter the transverse derivative being tested.

Direct integration of Eq.~\eqref{eq:linearfloweqs} gives
\begin{equation}
 M_\alpha^{\rm hyp}
 =\frac{2\lambda^{\alpha-1}}{\alpha}\sinh(\alpha\lambda\tau)
 \left(\abs q^\alpha+\abs p^\alpha\right).
 \label{eq:LDlinear}
\end{equation}
Crossing the stable manifold,
\begin{equation}
 \frac{\partial}{\partial q}\abs q^\alpha
 =\alpha\,\operatorname{sgn}(q)\abs q^{\alpha-1},\qquad q\ne0,
 \label{eq:LDqderiv}
\end{equation}
and the magnitude diverges as $q\to0$ because $\alpha-1<0$. The corresponding derivative of $\abs p^\alpha$ with respect to the momentum $p$ diverges as $p\to0$ when the unstable manifold is crossed. This is the established local nondifferentiability mechanism for this class of LDs \cite{LopesinoEtAl2017,NaikGarciaGarridoWiggins2019}.

The contrast with Eq.~\eqref{eq:LBlinear} is therefore explicit. Equation~\eqref{eq:LBlinear} has no spatial dependence whatsoever, whereas Eq.~\eqref{eq:LDlinear} contains the factors $\abs q^\alpha$ and $\abs p^\alpha$, whose transverse derivatives become singular on the invariant manifolds.

\subsection{Nonlinear corrections}

We now allow nonlinear terms in the Hamiltonian. The purpose is to determine how spatial variation first enters LB when the quadratic saddle approximation is no longer exact. Let
\begin{equation}
 z=(q,p,x,y)^{\mathsf T}
\end{equation}
denote displacement from the saddle equilibrium. The linear equations above can be written compactly as $\dot z=Az$, with
\begin{equation}
 A=\begin{pmatrix}
 \lambda&0&0&0\\
 0&-\lambda&0&0\\
 0&0&0&\omega\\
 0&0&-\omega&0
 \end{pmatrix}.
 \label{eq:Amatrix}
\end{equation}
The first two rows generate the expansion and contraction in the hyperbolic plane; the lower block generates the rotation in the center plane.

For the nonlinear system write
\begin{equation}
 \dot z=Az+R(z),
 \qquad R(0)=0,\quad DR(0)=0.
 \label{eq:localvf}
\end{equation}
The vector $R(z)$ contains everything omitted by the linear approximation. Because $A$ is the complete linearization, $DR(0)=0$. Suppose that inside a fixed local region
\begin{equation}
 \norm{DR(z)}\le K\norm z^{\nu},
 \qquad K>0,\quad \nu>0.
 \label{eq:DRbound}
\end{equation}
The derivative $DR(z)$ measures how much the local tangent dynamics differs from the constant matrix $A$; the exponent $\nu$ specifies how rapidly that difference vanishes as the saddle equilibrium is approached.

Let the initial condition lie within distance $\delta$ of the equilibrium. We consider intervals for which both the forward and backward trajectories remain in the chosen local region and satisfy the hyperbolic bound
\begin{equation}
 \norm{\Phi^{\pm t}(z_0)}\le C_0\delta e^{\lambda t},
 \label{eq:orbitbound}
\end{equation}
with $C_0$ independent of $\delta$. Applying variation of constants to the tangent equation and then carrying that estimate through the LB average gives
\begin{equation}
 \frac{\LB(z_0;\tau)}{e^{\lambda\tau}}-1
 =\OO\!\left[
 \frac{\delta^\nu(e^{\nu\lambda\tau}-1)}{\nu\lambda\tau}
 \right]
 +\OO\!\left(\delta^{2\nu}e^{\nu\lambda\tau}\right).
 \label{eq:LBbound}
\end{equation}
Appendix~\ref{app:proof} gives the derivation. The denominator is the spatially constant linear value in Eq.~\eqref{eq:LBlinear}. Thus Eq.~\eqref{eq:LBbound} measures the relative nonlinear departure from that value. For fixed $\tau$, the departure tends to zero as $\delta\to0$.

\subsection{The time required to leave the local saddle region}

The previous statement fixes the observation time. We next ask whether it remains true when the observation time grows while the initial neighborhood shrinks. Choose a small radius $\rho$ independent of $\delta$ within the local region. Equating the growth scale $C_0\delta e^{\lambda t}$ to $\rho$ defines the local hyperbolic escape-time scale
\begin{equation}
 \tau_\rho(\delta)
 =\frac{1}{\lambda}\ln\frac{\rho}{C_0\delta}
 =\frac{1}{\lambda}\ln\frac{1}{\delta}+\OO(1).
 \label{eq:escapetime}
\end{equation}
This is not the exit time of every trajectory: the equilibrium never leaves, and a point on a stable manifold need not leave in forward time. We use this scale only while both histories satisfy Eq.~\eqref{eq:orbitbound}.
Substitution into Eq.~\eqref{eq:LBbound} yields
\begin{equation}
 \frac{\LB(z_0;\tau_\rho)}{e^{\lambda\tau_\rho}}-1
 =\OO\!\left(\frac{1}{\ln(1/\delta)}\right)
 \longrightarrow0,
 \qquad \delta\to0.
 \label{eq:escapeflat}
\end{equation}
Thus the relative LB variation on the shrinking set of initial conditions tends to zero even on this logarithmically growing time scale, while both histories remain under local control.

This is one limiting procedure. The complementary experiment holds spatial resolution fixed while increasing the observation time. Neighboring trajectories can then accumulate increasingly different nonlinear histories after leaving the immediate saddle region, so spatial variation of LB can become concentrated near manifold-organized transport without producing an LD-type loss of differentiability. The next section separates these two effects.

\section{Model calculations: spatial scale, observation time, and coupling}
\label{sec:benchmarks}

The preceding section describes what happens as initial conditions approach the saddle equilibrium. We now examine how that result appears in calculations performed on neighborhoods of nonzero size and how it competes with increasing observation time. We first use a separable quartic Hamiltonian whose invariant manifolds are known analytically. We then introduce coupling between the hyperbolic and center degrees of freedom to determine whether the local flattening persists when separability is lost.

\subsection{Quartic saddle: spatial resolution and observation time}

Consider
\begin{equation}
 H=H_{\rm hyp}(x,p_x)+H_{\rm cen}(y,p_y),
 \label{eq:quarticsep}
\end{equation}
with
\begin{equation}
 H_{\rm hyp}=\frac{p_x^2}{2}-\frac{x^2}{2}+\frac{x^4}{4},
 \qquad
 H_{\rm cen}=\frac{p_y^2+y^2}{2}.
 \label{eq:quarticH}
\end{equation}
The pair $(x,p_x)$ is the hyperbolic degree of freedom; $(y,p_y)$ is the center degree of freedom. Hamilton's equations are
\begin{equation}
 \dot x=p_x,\qquad \dot p_x=x-x^3,
 \qquad
 \dot y=p_y,\qquad \dot p_y=-y.
 \label{eq:quarticeqs}
\end{equation}
Linearizing the hyperbolic equations at $x=p_x=0$ gives eigenvalues $+1$ and $-1$. Hence the local hyperbolic growth rate is $\lambda=1$, and the spatially constant linear result is $\LB=e^\tau$.

At energy $E=0.5$, setting $x=p_x=0$ leaves
\begin{equation}
 p_y^2+y^2=1,
\end{equation}
so the NHIM on this energy surface is the periodic orbit in the center plane defined by this circle. To display the surrounding invariant-manifold geometry in two dimensions, we intersect the three-dimensional energy surface with $y=0$ and retain crossings with $p_y>0$. Energy conservation then determines
\begin{equation}
 p_y=\left[2\left(E-H_{\rm hyp}(x,p_x)\right)\right]^{1/2}.
 \label{eq:quarticsection}
\end{equation}
Each allowed pair $(x,p_x)$ therefore specifies one point on the section, and the periodic orbit intersects it at $(0,0)$.

Because the two degrees of freedom are separable, $H_{\rm hyp}$ is independently conserved. The stable and unstable manifolds of the periodic orbit correspond in the hyperbolic subsystem to the separatrix energy $H_{\rm hyp}=0$, giving
\begin{equation}
 p_x=\pm x\sqrt{1-\frac{x^2}{2}}.
 \label{eq:quarticmanifolds}
\end{equation}
Near the origin the two branches reduce to $p_x\simeq\pm x$, the linear unstable and stable directions.

As in the LD calculation of Sec.~\ref{sec:localtheory}, the center degree of freedom cannot simply be omitted without explanation. Here the reason is different. The full tangent map is block diagonal,
\begin{equation}
 D\Phi^t=\begin{pmatrix}M_{\rm hyp}(t)&0\\0&R_t\end{pmatrix},
 \label{eq:quarticblock}
\end{equation}
where $R_t$ is an orthogonal rotation. Its singular values are exactly one. The hyperbolic block is symplectic, so its largest singular value is at least one. Consequently
\begin{equation}
 \sigma_{\max}(D\Phi^t)=\sigma_{\max}(M_{\rm hyp}(t)).
\end{equation}
Thus all spatial variation of LB on this separable section comes from the nonlinear hyperbolic motion.

We first ask how that spatial variation enters close to the periodic-orbit intersection. Let the initial point be $(x_0,p_{x0})$ and define $r^2=x_0^2+p_{x0}^2$. The expansion is taken for $r\ll1$ at fixed observation time $\tau$. The hyperbolic coordinate satisfies
\begin{equation}
 \ddot x=x-x^3.
\end{equation}
To leading order,
\begin{equation}
 x_{\rm L}(t)=x_0\cosh t+p_{x0}\sinh t,
\end{equation}
and $x(t)=x_{\rm L}(t)+\OO(r^3)$. A tangent perturbation $\xi$ obeys
\begin{equation}
 \ddot\xi=\left[1-3x(t)^2\right]\xi.
 \label{eq:quarticvar}
\end{equation}
Equivalently,
\begin{equation}
 \dot M_{\rm hyp}=[A_0+\Delta A(t)]M_{\rm hyp},
 \qquad
 A_0=\begin{pmatrix}0&1\\1&0\end{pmatrix},
\end{equation}
with
\begin{equation}
 \Delta A(t)=\begin{pmatrix}0&0\\-3x_{\rm L}(t)^2&0\end{pmatrix}+\OO(r^4).
\end{equation}
Variation of constants gives
\begin{equation}
 M_{\rm hyp}(t)=e^{A_0t}
 +\int_0^t e^{A_0(t-u)}\Delta A(u)e^{A_0u}\,\dd u
 +\OO(r^4).
 \label{eq:quarticMexp}
\end{equation}
Expanding the forward and backward largest singular values in Eq.~\eqref{eq:LB} then gives
\begin{equation}
 \frac{\LB(x_0,p_{x0};\tau)}{e^\tau}
 =1-c_x(\tau)x_0^2-c_p(\tau)p_{x0}^2+\OO(r^4),
 \label{eq:quarticexpansionxp}
\end{equation}
where
\begin{align}
 c_x(\tau)&=\frac{3[\cosh(2\tau)-1]}{8\tau}+\frac{3\tau}{4},\\
 c_p(\tau)&=\frac{3[\cosh(2\tau)-1]}{8\tau}-\frac{3\tau}{4}.
 \label{eq:cxcp}
\end{align}
Both coefficients are positive for $\tau>0$. The first spatial dependence of LB is therefore smooth and quadratic. The separatrices in Eq.~\eqref{eq:quarticmanifolds} are not singular sets of the LB field. Local expansions around hyperbolic structures also play a central role in analytical studies of LDs \cite{LopesinoEtAl2017,NaikGarciaGarridoWiggins2019}; the mechanism here is different because LB is constructed from tangent-map singular values rather than from a trajectory integral.

\begin{figure}[ht]
\centering
\includegraphics[width=0.98\linewidth]{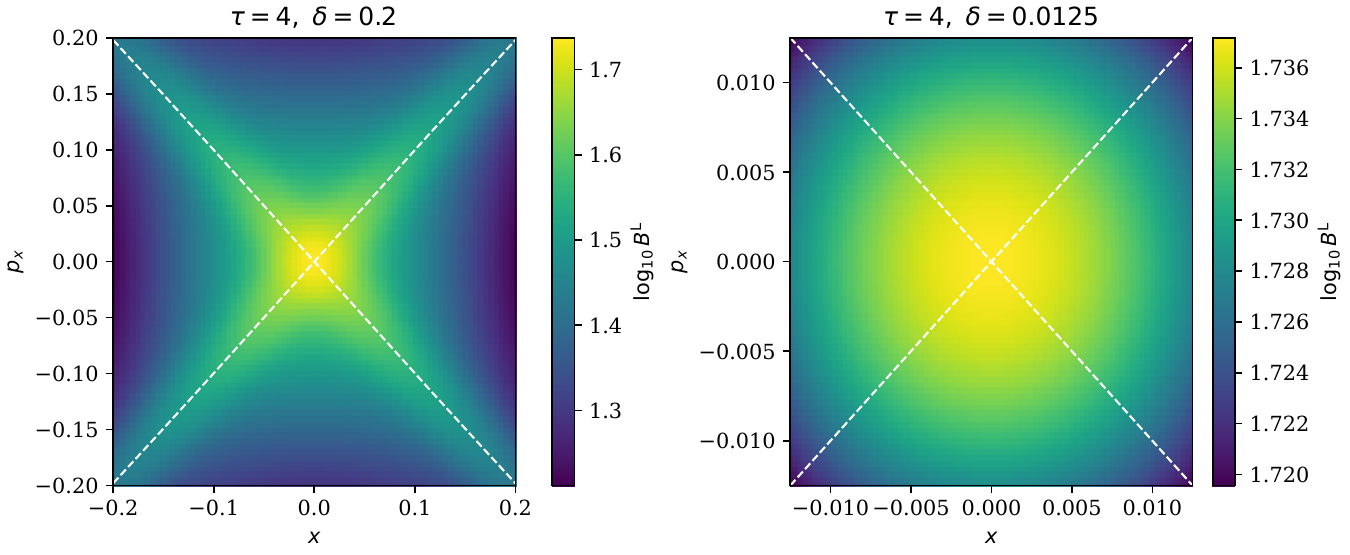}
\caption{Quartic model at $E=0.5$ and fixed observation time $\tau=4$. Left: $|x|,|p_x|\le0.2$. Right: $|x|,|p_x|\le0.0125$. The colorbars show $\log_{10}B^{\mathrm L}$, with a separate color range in each panel. The dashed curves are the exact stable and unstable separatrices from Eq.~\eqref{eq:quarticmanifolds}. Shrinking the spatial window reduces the relative LB variation rather than producing a singular manifold signature.}
\label{fig:quarticfields}
\end{figure}

Figure~\ref{fig:quarticfields} shows the first limiting experiment. The observation time is fixed at $\tau=4$ while the spatial window is reduced. A pronounced separatrix-organized pattern is visible on the larger window, whereas on the smaller window the relative variation is much weaker and approaches the smooth quadratic form of Eq.~\eqref{eq:quarticexpansionxp}.

\begin{figure}[ht]
\centering
\includegraphics[width=0.98\linewidth]{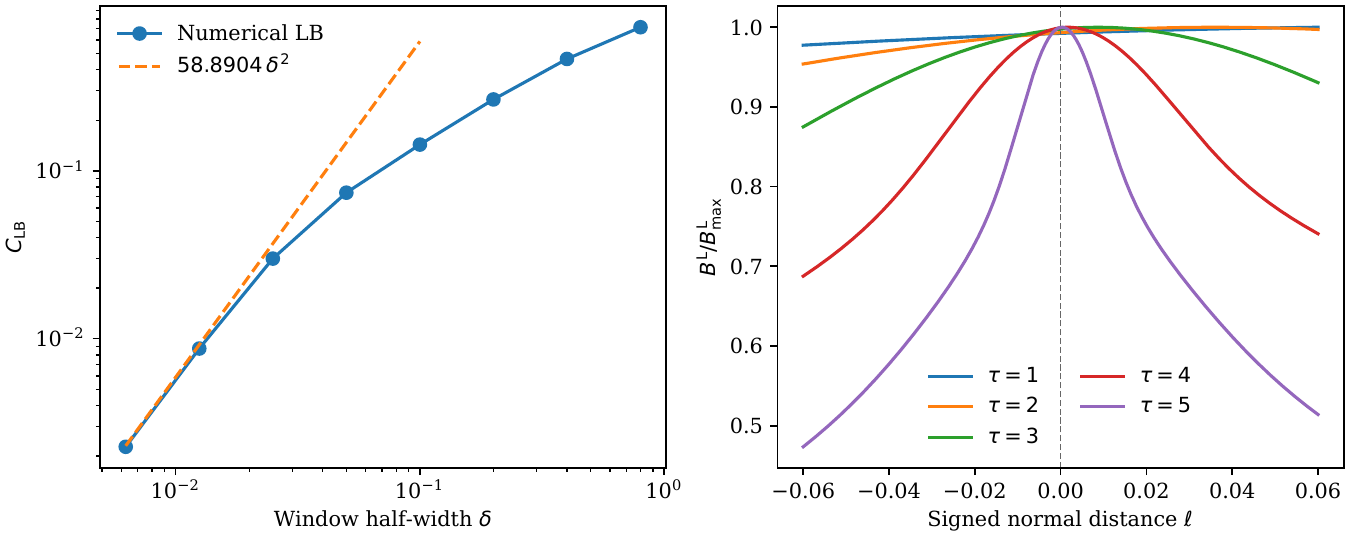}
\caption{Two complementary limits in the quartic model. Left: at fixed $\tau=4$, the relative LB variation $C_{\rm LB}$ decreases with the window half-width $\delta$; the dashed guide is the continuum leading term $58.8904\,\delta^2$. Right: LB on the normal transect through the unstable separatrix at $x_*=0.09$, Eq.~\eqref{eq:transectbase}. The horizontal coordinate is signed distance $\ell$ along the transect, and the dashed vertical line marks $\ell=0$. Each curve is divided by its own maximum $B^{\mathrm L}_{\max}$ over the displayed interval, so the panel compares spatial profiles rather than their overall magnitudes. Increasing $\tau$ concentrates the spatial variation near the separatrix without a loss of differentiability there.}
\label{fig:quarticlimits}
\end{figure}

To quantify this flattening, let
\begin{equation}
 \langle F\rangle_\delta
 =\frac{1}{4\delta^2}\int_{-\delta}^{\delta}\int_{-\delta}^{\delta}
 F(x,p_x)\,\dd x\,\dd p_x
 \label{eq:spatialaverage}
\end{equation}
be the uniform area average over the square $|x|,|p_x|\le\delta$, and define
\begin{equation}
 C_{\rm LB}(\delta)
 =\frac{
 \left\langle(\LB-\langle\LB\rangle_\delta)^2\right\rangle_\delta^{1/2}}
 {\langle\LB\rangle_\delta}.
 \label{eq:cvdef}
\end{equation}
For a uniform square, $\langle x^2\rangle=\langle p_x^2\rangle=\delta^2/3$ and $\operatorname{var}(x^2)=\operatorname{var}(p_x^2)=4\delta^4/45$. Equation~\eqref{eq:quarticexpansionxp} therefore gives
\begin{equation}
 C_{\rm LB}(\delta)
 =\frac{2}{\sqrt{45}}\sqrt{c_x^2+c_p^2}\,\delta^2+\OO(\delta^4).
 \label{eq:quarticcvanalytic}
\end{equation}
At $\tau=4$, $c_x=142.6387$ and $c_p=136.6387$, so the continuum coefficient is $58.8904$. For Fig.~\ref{fig:quarticlimits}, we approximate the uniform area average with tensor-product composite Simpson quadrature on an $81\times81$ Cartesian grid. Fourth-order Runge--Kutta integration uses time step $0.005$, and composite Simpson quadrature of the split-time integral uses spacing $0.01$. At the smallest window, $\delta=0.00625$, this gives $C_{\rm LB}=0.00227022$, or $C_{\rm LB}/\delta^2=58.12$. Applying the same spatial quadrature to the quadratic field gives $C_{\rm LB}=0.00230880$; the full nonlinear value is about $1.7\%$ smaller. Halving both time spacings changes the calculated contrasts by less than $9\times10^{-10}$ relative. Increasing the spatial grid to $161\times161$ changes them by less than $4\times10^{-6}$ relative.

The complementary experiment keeps the spatial transect fixed and increases $\tau$. Figure~\ref{fig:quarticlimits} shows both results. The left panel displays the shrinking-window decrease of $C_{\rm LB}$ at $\tau=4$. The right panel follows LB along a transect normal to the exact unstable separatrix at
\begin{equation}
 x_*=0.09,\qquad p_{x*}=x_*\sqrt{1-x_*^2/2}.
 \label{eq:transectbase}
\end{equation}
Let $\ell$ denote signed distance along this normal transect, with $\ell=0$ at the separatrix and positive $\ell$ toward increasing $p_x$. As $\tau$ increases, the spatial variation becomes increasingly concentrated near the separatrix. At $\tau=3$ the numerical maximum is displaced by $0.0087$ from the separatrix. Writing Eq.~\eqref{eq:quarticexpansionxp} in the local linear stable/unstable coordinates $U=(x+p_x)/\sqrt2$ and $S=(x-p_x)/\sqrt2$ produces a $US$ cross term; evaluated at the base point in Eq.~\eqref{eq:transectbase}, the same local expansion predicts an offset $0.0114$. The base point is already near the edge of the quantitative range of the quadratic expansion, so the roughly $25\%$ difference between these two offsets is not unexpected. Over $\tau=3,4,5$, the local curvature width $w_c=[\LB/|\partial_\ell^2\LB|]^{1/2}$ at the maximum scales as $e^{-0.97\tau}$, close to the $e^{-\lambda\tau}$ local resolution scale with $\lambda=1$. These quantitative checks connect the fixed-resolution transect to the local expansion while preserving the important point: the LB field remains smooth across the separatrix.

\subsection{Coupling the hyperbolic and center degrees of freedom}
\label{sec:floquetcheck}

The calculations above are straightforward because the hyperbolic and center degrees of freedom are separable. We now ask whether the local flattening of LB persists when these motions are coupled. To isolate the effect of the coupling, consider
\begin{equation}
 H=\frac12(p_x^2+p_y^2)-\frac{x^2}{2}+\frac{x^4}{4}+\frac{y^2}{2}
 +\epsilon x^2y+\eta x^3y,
 \label{eq:coupledH}
\end{equation}
with $\epsilon=0.15$, $\eta=0.05$, and $E=0.5$. The last two terms couple the hyperbolic and center degrees of freedom. Because they vanish when $x=0$, the motion
\begin{equation}
 x=p_x=0,\qquad y(t)=\cos t,\qquad p_y(t)=-\sin t
 \label{eq:coupledPO}
\end{equation}
remains an exact periodic orbit of period $T=2\pi$.

Before using this orbit as the reference structure for the LB calculation, we must establish that it remains hyperbolic after the coupling is introduced and determine its transverse growth rate. We therefore ask what happens to an infinitesimal displacement away from the orbit in the $(x,p_x)$ directions. Let $\xi(t)$ be a small displacement in $x$ and let $\pi(t)$ be its conjugate momentum displacement. Linearization about Eq.~\eqref{eq:coupledPO} gives
\begin{equation}
 \dot\xi=\pi,\qquad
 \dot\pi=[1-2\epsilon\cos t]\xi,
\end{equation}
and hence
\begin{equation}
 \ddot\xi=[1-2\epsilon\cos t]\xi.
 \label{eq:coupledvar}
\end{equation}
The term proportional to $\eta x^3y$ does not enter this linear equation because its contribution to the derivative of the $x$ equation vanishes at $x=0$. It affects the nonlinear dynamics away from the orbit but not its linear stability.

The coefficient in Eq.~\eqref{eq:coupledvar} is periodic with the same period $T$ as the orbit. Integrating the linearized equations over one period defines a $2\times2$ tangent map $M_\perp(T)$ that takes an initial transverse displacement to the displacement one period later. Its eigenvalues are the Floquet multipliers,
\begin{equation}
 \mu_s=1.92\times10^{-3},\qquad
 \mu_u=5.20\times10^2.
\end{equation}
One multiplier contracts and the other expands, confirming that the periodic orbit is hyperbolic. The corresponding exponential growth rate per unit time is
\begin{equation}
 \lambda_F=\frac{1}{T}\ln\mu_u=0.995.
 \label{eq:floquetlambda}
\end{equation}
This plays the same role for the periodic orbit that $\lambda=1$ played for the local saddle dynamics in the separable model.
We now test whether LB becomes locally uniform as initial conditions approach this periodic orbit while the energy is held fixed. Nearby initial conditions must be compared at the same phase of the periodic motion. We therefore use the section $p_y=0$ and vary $(x,p_x)$. For each pair $(x,p_x)$ near the orbit, the positive solution for $y$ is obtained from
\begin{equation}
 \frac{y^2}{2}+(\epsilon x^2+\eta x^3)y
 +H_{\rm hyp}(x,p_x)-E=0,
 \label{eq:coupledsection}
\end{equation}
so that every point lies on the fixed energy surface. At the periodic orbit, $x=p_x=0$ and Eq.~\eqref{eq:coupledsection} gives $y=1$. We then compute $C_{\rm LB}$ on progressively smaller square parameter windows $|x|,|p_x|\le\delta$ around that section point. The area averages use tensor-product Simpson weights on a $41\times41$ grid. The observation intervals end at $\tau=\pi$ and $2\pi$, with window sequences $\delta=0.04/2^k$, $k=0,\ldots,9$ and $k=0,\ldots,8$, respectively.

\begin{figure}[ht]
\centering
\includegraphics[width=0.70\linewidth]{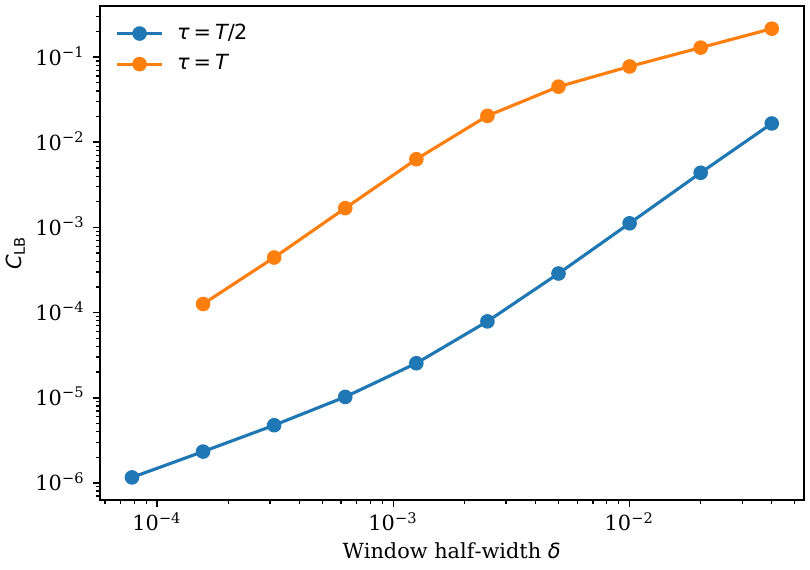}
\caption{Shrinking-window LB variation for the coupled model at $E=0.5$. Initial conditions lie on the fixed-energy section defined by Eq.~\eqref{eq:coupledsection}. Here $T=2\pi$ is the periodic-orbit period and $\tau$ is the observation time. For both $\tau=T/2$ and $\tau=T$, the relative spatial variation $C_{\rm LB}$ tends to zero as the section window around the hyperbolic periodic orbit is reduced.}
\label{fig:coupledscaling}
\end{figure}

Figure~\ref{fig:coupledscaling} shows the result for observation times $T/2$ and $T$. In both cases
\begin{equation}
 C_{\rm LB}\longrightarrow0\qquad\text{as}\qquad\delta\longrightarrow0.
\end{equation}
Thus the local flattening found in the separable quartic model is not a consequence of separability; it persists when the hyperbolic and center degrees of freedom are coupled.

The changing slope of the $T/2$ curve on the logarithmic plot also has a simple local symmetry origin. The decisive point is symmetry, not merely the appearance of tangent-matrix entries that are linear in the transverse displacement: when $\eta=0$, the Hamiltonian is invariant under $(x,p_x)\mapsto(-x,-p_x)$, so the LB field on the section is even and an $O(\delta)$ spatial variation is forbidden, even though the $\epsilon x^2y$ term produces tangent-matrix entries linear in $x$. The $\eta x^3y$ term breaks this symmetry; its contribution $-3\eta x^2\cos t$ to the transverse equation then permits an odd $O(\delta)$ LB variation. At moderately small windows the even quadratic contributions still dominate and the apparent slope is near two; only at the smallest windows does the $O(\delta)$ term dominate and the local slope approach one.

To compare these contributions, we separate the LB field into its odd and even parts under $(x,p_x)\mapsto(-x,-p_x)$. Their spatial standard deviations, normalized by the mean of the full LB field and divided by $\delta$ and $\delta^2$, respectively, estimate the local linear and quadratic coefficients. At the smallest sampled windows, increasing the observation interval from $T/2$ to $T$ increases these coefficients by factors of about $30$ and $390$. The quadratic contribution therefore grows faster over this interval. The ratio of the coefficients gives a crossover scale of $1.33\times10^{-3}$ at $T/2$ and $1.01\times10^{-4}$ at $T$. The latter lies below the smallest full-period window, $\delta=1.5625\times10^{-4}$, explaining why that sequence remains close to quadratic at its smallest resolved scales. The fitted slope over any one finite range is not the conclusion of the calculation; $C_{\rm LB}\to0$ is.

\section{HCN/CNH: molecular model, dividing surfaces, and gap times}
\label{sec:hcnmodel}

\subsection{Hamiltonian and saddle geometry}

We now turn to the planar, zero-total-angular-momentum, three-degree-of-freedom HCN/CNH model. The NHIMs, stable and unstable manifolds, and dividing surfaces governing isomerization in this model were constructed with Poincaré--Birkhoff normal forms in Ref.~\cite{WaalkensBurbanksWiggins2004}. Mean passage times were subsequently computed in Ref.~\cite{WaalkensBurbanksWiggins2005}, and gap-time and lifetime distributions were analyzed in detail in Ref.~\cite{EzraWaalkensWiggins2009}. We use the same Murrell--Carter--Halonen (MCH) ground-state potential \cite{MurrellCarterHalonen1982}. We retain the notation CNH used in that phase-space literature; the same linear isomer is more commonly denoted HNC in spectroscopy.

Let $r_{\rm CN}$ be the C--N separation and $R$ the distance from H to the center of mass of CN. With the corresponding reduced masses $\mu_{\rm CN}$ and $\mu_{\rm H,CN}$, define the mass-scaled Jacobi coordinates
\begin{equation}
 q_1=\sqrt{\mu_{\rm CN}}\,r_{\rm CN},\qquad
 q_2=\sqrt{\mu_{\rm H,CN}}\,R,
\end{equation}
with bend angle $\gamma$ and conjugate momenta $p_1,p_2,p_\gamma$. The Hamiltonian is
\begin{equation}
 H=\frac12(p_1^2+p_2^2)
 +\frac12\left(\frac{1}{q_1^2}+\frac{1}{q_2^2}\right)p_\gamma^2
 +V_{\rm MCH}\!\left(
 \frac{q_1}{\sqrt{\mu_{\rm CN}}},
 \frac{q_2}{\sqrt{\mu_{\rm H,CN}}},\gamma\right).
 \label{eq:HCNH}
\end{equation}
We use the same \AA--amu--eV mechanical units as the earlier HCN/CNH phase-space calculations \cite{WaalkensBurbanksWiggins2004}. With this choice the corresponding unit of time is
\begin{equation}
 t_{\rm u}=\sqrt{\frac{{\rm amu}\,\text{\AA}^2}{{\rm eV}}}
 =0.0101805\ {\rm ps}.
 \label{eq:timeunit}
\end{equation}
We solve the stationary-point equations $\nabla V_{\rm MCH}=0$ directly in our implementation of the potential. This gives
\begin{equation}
 (r_{\rm CN},R,\gamma)
 =(1.1394\ \text{\AA},1.2084\ \text{\AA},1.16823\ {\rm rad})
\end{equation}
and
\begin{equation}
 E_s=-12.0827\ {\rm eV}.
\end{equation}
These values agree with the saddle used in the earlier phase-space calculation of Ref.~\cite{WaalkensBurbanksWiggins2004}. Linearizing Hamilton's equations at this equilibrium gives
\begin{equation}
 \pm2.16941,\qquad \pm i\,5.99984,\qquad \pm i\,4.26275.
 \label{eq:HCNspectrum}
\end{equation}
Thus there is one hyperbolic degree of freedom and two center degrees of freedom.

The reactive calculations below use
\begin{equation}
 E=E_s+0.2\ {\rm eV}.
 \label{eq:HCNenergy}
\end{equation}
This is the energy used for the phase-space structures in Ref.~\cite{WaalkensBurbanksWiggins2004} and for the subsequent passage-time and gap-time calculations in Refs.~\cite{WaalkensBurbanksWiggins2005,EzraWaalkensWiggins2009}. The saddle equilibrium itself lies at $E_s$; on the higher energy surface in Eq.~\eqref{eq:HCNenergy}, the transition-state geometry is organized by a three-dimensional NHIM and its associated dividing surface.

\subsection{How the incoming reactive ensemble is constructed}

To construct and sample the dividing surface we use local canonical coordinates obtained from a tenth-order Poincaré--Birkhoff normal form, following the construction of Refs.~\cite{WaalkensBurbanksWiggins2004,WaalkensSchubertWiggins2008}. The normal form separates the hyperbolic degree of freedom from the two elliptic degrees of freedom near the saddle region and supplies the direct and inverse transformations between the local normal-form variables and the physical Jacobi variables.

We use \emph{center degrees of freedom} when discussing the local saddle geometry. In the normal-form reaction-dynamics literature, the same elliptic degrees of freedom are conventionally called \emph{bath modes}; we use that terminology when discussing their normal-form actions and angles.

The incoming ensemble is constructed with the same dividing-surface flux measure and normal-form sampling procedure used in Refs.~\cite{WaalkensBurbanksWiggins2005,EzraWaalkensWiggins2009}. We use a fixed ensemble of $25\,000$ initial conditions sampled uniformly with respect to the natural incoming flux measure on one incoming hemisphere. In action-angle variables this is equivalent to uniform density over the allowed bath-action region together with uniform bath angles. The hyperbolic degree of freedom is described by the positive reactive action $I$, and on the incoming hemisphere its normal-form coordinates satisfy
\begin{equation}
 \bar q_1=\bar p_1=\sqrt I.
 \label{eq:incomingI}
\end{equation}
Thus small $I$ places the initial condition close to the NHIM in the hyperbolic directions. Mapping the ensemble back to physical coordinates gives an rms energy-placement error of $5.0\times10^{-7}$ eV. This is the discrepancy between the physical Hamiltonian and the target energy at the initial points; it is distinct from the energy drift during integration. The sampled actions and angles are retained in the reproducibility package, together with the reconstructed physical initial conditions \cite{WigginsRepro2026}.

Two reflection-related NHIM-anchored dividing surfaces bound the HCN well at this energy. We denote the incoming and outgoing hemispheres of the first by $DS_{1,{\rm in}}$ and $DS_{1,{\rm out}}$, and the outgoing hemisphere of the second by $DS_{2,{\rm out}}$. Every trajectory enters through $DS_{1,{\rm in}}$. We call an orbit channel 1 if it later leaves through $DS_{1,{\rm out}}$ and channel 2 if it leaves through $DS_{2,{\rm out}}$, following Ref.~\cite{EzraWaalkensWiggins2009}.

\subsection{Locating an exit}

A gap time ends when the trajectory first crosses either outgoing hemisphere. The crossing must be identified in the local chart where the normal-form dividing surface is defined. Evaluating its polynomial equation far from that chart can produce roots that do not represent a passage through the transition state.

For each saddle, the degree-ten inverse normal-form transformation gives the local hyperbolic coordinates $\bar q_1(z)$ and $\bar p_1(z)$ of a physical state $z$. We locate roots of
\begin{equation}
 g(z)=\bar q_1(z)-\bar p_1(z)=0
 \label{eq:exitroot}
\end{equation}
and accept an outgoing crossing only when $\bar q_1<0$, $\bar p_1<0$, and $\dot g<0$. We also restrict the crossing to the local chart. Let $S$ be the real linear canonical transformation from the saddle normal-form coordinates to physical displacements, and let $z^\ddagger$ be the saddle point in that chart. The accepted root must satisfy $\|S^{-1}(z-z^\ddagger)\|_2\le0.6$. For the second saddle chart the reflection changes the signs of both $\gamma$ and $p_\gamma$. These conditions are applied to the continuous root after it has been located; the locality condition does not introduce a discontinuity into the root function.

We propagate every incoming point from time zero until its first accepted outgoing crossing, using DOP853, an adaptive Runge--Kutta method that adjusts its step to control local numerical error \cite{HairerNorsettWanner1993}. A second calculation repeats the entire ensemble with tighter tolerances and a smaller maximum step. Both calculations reach an observed exit for every trajectory. We call the first calculation primary and use its gap times and channels for the results below; the refined calculation measures their numerical sensitivity. Appendix~\ref{app:numerics} gives the settings and comparison.

The primary calculation gives 2533 channel-1 and 22467 channel-2 exits, so $N_2/N_1=8.87$, compared with the published value $8.76$ \cite{EzraWaalkensWiggins2009}. The refined calculation changes the channel of 94 trajectories and leaves the branching fractions close to their primary values. The longest gaps have a much stronger effect on mean gap times. Their sensitivity is examined separately in Appendix~\ref{app:numerics}; completion of every trajectory does not by itself establish the accuracy of every late exit.

For the LB calculations we propagate each physical trajectory together with the full $6\times6$ variational equations. Unless stated otherwise, singular values are computed with the Euclidean norm on the physical mass-scaled canonical coordinates $(q_1,q_2,\gamma,p_1,p_2,p_\gamma)$. Numerical checks are collected in Appendix~\ref{app:numerics}.

\section{Local behavior near the HCN/CNH saddle equilibrium}
\label{sec:hcnlocal}

Section~\ref{sec:localtheory} derived a local result near a saddle equilibrium without restricting the calculation to a single energy surface. Before turning to the reactive ensemble on $E=E_s+0.2$ eV, we test that result using the full HCN/CNH Hamiltonian near its saddle equilibrium at $E_s$. This is not a calculation on the three-dimensional NHIM that exists on the higher energy surface.

The molecular phase space is six-dimensional. Let $z^\ddagger$ denote the saddle equilibrium and let $e_u$ and $e_s$ be Euclidean-normalized unstable and stable eigenvectors of the linearized Hamiltonian equations. We examine the two-dimensional affine slice
\begin{equation}
 z=z^\ddagger+a e_u+b e_s,
 \qquad |a|,|b|\le\delta.
 \label{eq:hcnplane}
\end{equation}
Equation~\eqref{eq:hcnplane} maps the square $|a|,|b|\le\delta$ in parameter space into the plane spanned by the two hyperbolic eigendirections of the full six-dimensional phase space. It is not contained in the five-dimensional energy surface $E=E_s+0.2$ eV; the energy varies across the slice. Reducing $\delta$ therefore approaches the saddle equilibrium along its hyperbolic directions. It should not be interpreted as approaching the higher-energy NHIM.

We first ask whether the molecular LB field becomes locally uniform in this slice, as predicted by Eq.~\eqref{eq:LBbound}. At observation time $\tau=0.00509$ ps (the computational value $\tau/t_{\rm u}=0.5$), the dimensionless hyperbolic amplification exponent is $\lambda\tau/t_{\rm u}\simeq1.08$. Using the spatial contrast in Eq.~\eqref{eq:cvdef} with $(x,p_x)$ replaced by $(a,b)$ gives
\begin{equation}
 C_{\rm LB}\simeq0.118\,\delta^{1.00}.
 \label{eq:hcnlocalfit}
\end{equation}
The fit uses all six windows $\delta=0.04/2^k$, $k=0,\ldots,5$, with tensor-product Simpson quadrature on a $41\times41$ grid.
The observed linear scaling is consistent with the molecular Taylor expansion. A generic cubic Hamiltonian term produces a quadratic correction $R(z)=\OO(\norm z^2)$ to the vector field and therefore $DR(z)=\OO(\norm z)$. In Eq.~\eqref{eq:DRbound} this is $\nu=1$, so the local estimate permits $\OO(\delta)$ spatial variation. The quartic model gave $\OO(\delta^2)$ because its reflection symmetry removed the cubic Hamiltonian term.

The same behavior persists at longer observation times when the spatial window is reduced accordingly. At $\tau=0.0234$ ps (computational value $2.30$, with $\lambda\tau/t_{\rm u}\simeq4.99$), a fit over the three windows $\delta=1.5625\times10^{-4}/2^k$, $k=0,1,2$, gives $C_{\rm LB}\propto\delta^{1.0009}$. At $\tau=0.0328$ ps (computational value $3.22$, with $\lambda\tau/t_{\rm u}\simeq6.99$), the three windows $\delta=3.90625\times10^{-5}/2^k$ give $C_{\rm LB}\propto\delta^{1.0027}$. The molecular calculation therefore shows the same local flattening even after substantial exponential amplification.

\begin{figure}[ht]
\centering
\includegraphics[width=0.76\linewidth]{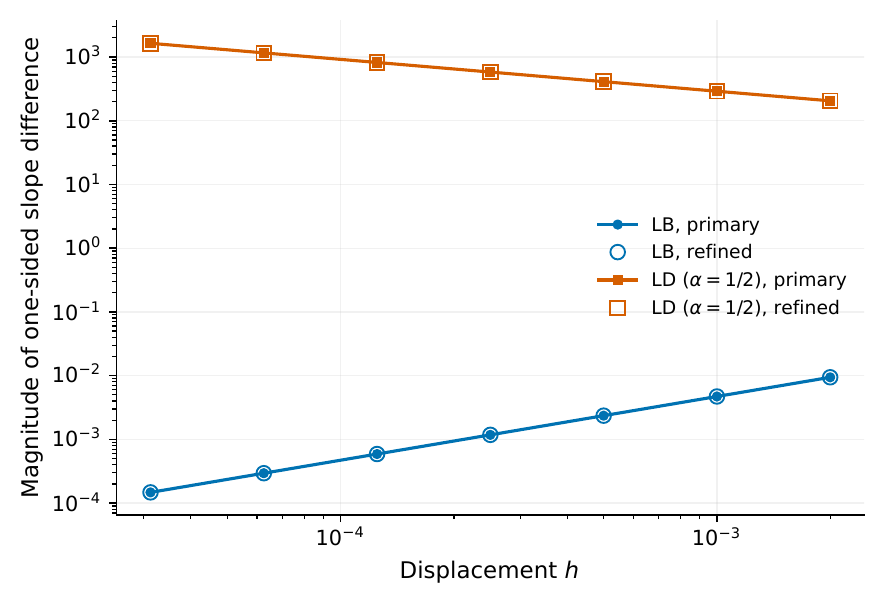}
\caption{HCN/CNH one-sided finite differences at observation time $\tau=0.00509$ ps ($\tau/t_{\rm u}=0.5$). The ordinate is $\lvert D_h^+G-D_h^-G\rvert$, with $D_h^\pm$ defined in Eq.~\eqref{eq:onesided}. Circles represent $G=B^{\mathrm L}$; squares represent $G=M_{1/2}$. Filled markers and connecting lines show the primary calculation; open markers show the refinement. The horizontal coordinate $h>0$ is the displacement along the unstable eigendirection through the saddle equilibrium. The points are calculated from the full molecular equations and, for LB, their variational equations; details and convergence checks are given in Appendix~\ref{app:numerics}. Fits to the primary values over the seven displacements $h=0.002/2^k$, $k=0,\ldots,6$, give slopes $1.00001$ for LB and $-0.500000$ for the LD. This slice is not constrained to a fixed energy and does not pass through the $E_s+0.2$ eV NHIM.}
\label{fig:hcndiff}
\end{figure}

We next compare differentiability of LB and the $\alpha=1/2$ LD. The LD result of Sec.~\ref{sec:localtheory} is naturally stated in local coordinates whose hyperbolic axes are aligned with the stable and unstable directions. For this comparison both diagnostics are evaluated at $\tau=0.00509$ ps ($\tau/t_{\rm u}=0.5$). The LD uses the velocity components in the physical mass-scaled coordinates, and LB uses the Euclidean norm in those same coordinates. Along the unstable eigendirection through the molecular equilibrium, define for either scalar field $G$ and positive displacement $h$
\begin{equation}
 D_h^+G=\frac{G(z^\ddagger+h e_u)-G(z^\ddagger)}{h},\qquad
 D_h^-G=\frac{G(z^\ddagger)-G(z^\ddagger-h e_u)}{h}.
 \label{eq:onesided}
\end{equation}
If $G$ is differentiable at the equilibrium, $D_h^+G-D_h^-G\to0$ as $h\to0$. For the $\alpha=1/2$ LD, Eq.~\eqref{eq:LDqderiv} instead gives the characteristic $h^{-1/2}$ divergence in the aligned hyperbolic coordinate. Figure~\ref{fig:hcndiff} shows that the LB one-sided slopes approach one another, whereas the LD exhibits the expected nonsmooth power law. The LD power law here also reflects that the velocity vanishes at the equilibrium; such behavior is not unique to hyperbolic directions. This calculation tests regularity along the chosen direction, not manifold selectivity away from the equilibrium.

The calculation also illustrates the norm dependence described in Sec.~\ref{sec:metric}. At the shortest observation time, the linear saddle value is $e^{\lambda\tau/t_{\rm u}}=2.96$ in orthonormal local saddle coordinates, whereas the same linearized molecular dynamics gives LB approximately $21.6$ when the Euclidean norm is imposed on the physical mass-scaled coordinates. The numerical LB value changes with the chosen norm; the decrease in spatial variation as the saddle equilibrium is approached does not.

\section{Pre-exit LB, reactive action, and gap times}
\label{sec:gaptimes}

For each incoming initial condition $z_i$ on $DS_{1,{\rm in}}$, $i=1,\ldots,N$ with $N=25\,000$, the calculation associates three numbers with the same trajectory. The incoming reactive action $I_i$ specifies how the trajectory is launched relative to the NHIM. The pre-exit Lagrangian betweenness
\begin{equation}
 B_i^{\rm L}=B^{\rm L}(z_i;\tau_*)
 \label{eq:Bi}
\end{equation}
is measured early in the motion. The gap time $t_{{\rm gap},i}$ is obtained only after the trajectory has subsequently reached an outgoing dividing surface. The purpose of this section is to determine how these three quantities are related and, in particular, what dynamical information is contained in the early LB value.

The shortest gap time in the ensemble is
\begin{equation}
 t_{\rm gap}^{\rm min}=0.02966\ {\rm ps}.
\end{equation}
We therefore choose
\begin{equation}
 \tau_*=0.0285\ {\rm ps},
 \label{eq:taustar}
\end{equation}
corresponding to the convenient computational duration $\tau_*/t_{\rm u}=2.8$. Because $\tau_*<t_{\rm gap}^{\rm min}$, every value $B_i^{\rm L}$ used below is determined before any trajectory in the ensemble has exited.

\subsection{Pre-exit LB and eventual gap time}

We first ask whether trajectories with larger LB before any exit occurs tend subsequently to have longer gap times. The comparison can be made without assuming any particular formula relating the two quantities. We sort the same $25\,000$ trajectories once from smallest to largest $B_i^{\rm L}$ and a second time from smallest to largest $t_{{\rm gap},i}$. We then ask how similar these two orderings are. If nearly the same trajectories occur in the same order, the comparison is close to $+1$; if the orderings are nearly reversed, it is close to $-1$; if there is no systematic relation between the two orderings, it is close to zero. This comparison is Spearman's rank correlation \cite{Spearman1904}. We denote the resulting single ensemble number by $\rho_S(B^{\rm L},t_{\rm gap})$. The symbols inside $\rho_S$ name the two quantities used to rank the full ensemble; they are not values at which a function is being evaluated.

For the present ensemble,
\begin{equation}
 \rho_S(B^{\rm L},t_{\rm gap})=0.909.
 \label{eq:LB909}
\end{equation}
Thus there is a strong tendency for the trajectories with relatively large pre-exit LB to be the same trajectories that have relatively long eventual gap times. Appendix~\ref{app:numerics} gives the corresponding comparisons with one-sided finite-time stretching measures and with the LD magnitude; those secondary comparisons are not needed for the mechanism developed here.

\begin{figure}[ht]
\centering
\includegraphics[width=0.75\linewidth]{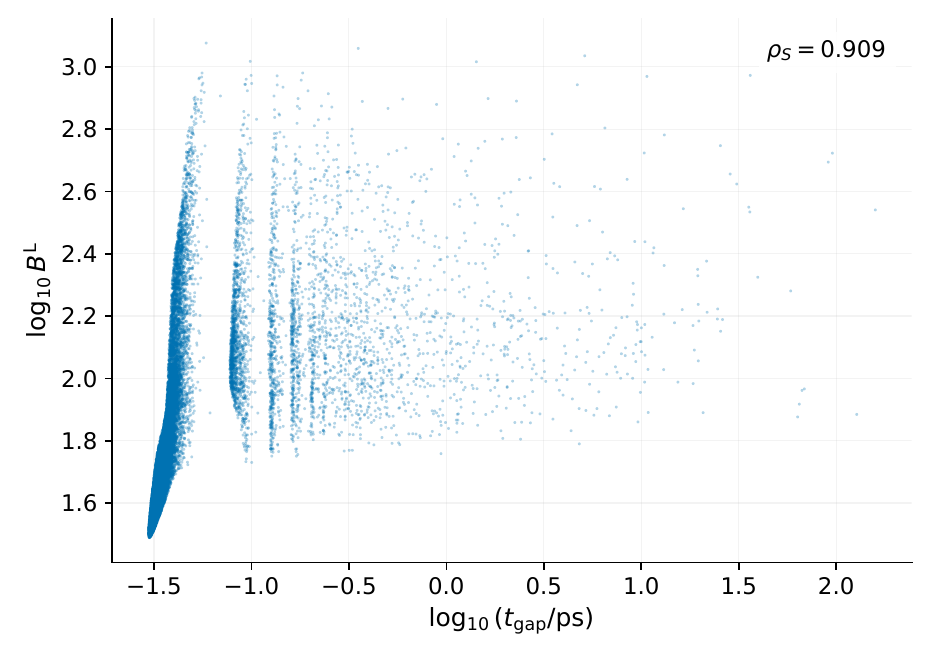}
\caption{Each point represents one of the $25\,000$ incoming trajectories. The horizontal coordinate is its eventual gap time and the vertical coordinate is its LB value measured earlier at $\tau_*=0.0285$ ps, before any trajectory has exited. Both coordinates are displayed as base-10 logarithms to show the dense short-passage population and the long gap-time tail on the same plot. The overall upward ordering is summarized by $\rho_S=0.909$; the scatter shows that LB does not determine the gap time through a single-valued formula.}
\label{fig:LBgap}
\end{figure}

Figure~\ref{fig:LBgap} shows the $25\,000$ trajectory pairs underlying Eq.~\eqref{eq:LB909}. Each point represents one initial condition: its horizontal position is the eventual gap time of that trajectory and its vertical position is the LB value measured earlier at $\tau_*$. The dense short-passage population rises overall from smaller LB and shorter gap times toward larger LB and longer gap times. The vertical bands reflect the recurrence structure of the short HCN/CNH gap-time distribution discussed in Ref.~\cite{EzraWaalkensWiggins2009}, while the much longer gap-time tail is more broadly scattered. The points therefore do not lie on a single curve. Equation~\eqref{eq:LB909} summarizes the much simpler statement that, across the ensemble, a trajectory ranked high by early LB is very likely also to rank high by eventual gap time.

The strong rank relation tells us that the two quantities are connected, but it does not yet tell us why. We therefore ask whether both are controlled by the same feature of the incoming initial condition.

\subsection{The incoming reactive action}

On the incoming dividing surface, Eq.~\eqref{eq:incomingI} gives $\bar q_1=\bar p_1=\sqrt I$. Thus $I$ measures the initial displacement from the NHIM in the hyperbolic directions: smaller $I$ means that the initial condition is chosen closer to the NHIM. We now use the same rank comparison as above, first between $I_i$ and $t_{{\rm gap},i}$ and then between $I_i$ and $B_i^{\rm L}$. For the full $25\,000$-trajectory ensemble,
\begin{equation}
 \rho_S(I,t_{\rm gap})=-0.921,
 \qquad
 \rho_S(I,B^{\rm L})=-0.976.
 \label{eq:Icorrelations}
\end{equation}
Here again each number is obtained by ranking the same $25\,000$ trajectories twice. The first negative value means that the two rankings are nearly reversed: trajectories with smaller $I$ tend strongly to have longer gap times. The second means that trajectories with smaller $I$ tend strongly to have larger pre-exit LB. These two comparisons suggest a common \mbox{dynamical} explanation in terms of the incoming reactive action. We next examine how the local passage near the NHIM can account for this dependence.

\subsection{Local hyperbolic mechanism}

To interpret these trends, consider the local passage near the higher-energy NHIM. Let $\Lambda$ approximate the transverse hyperbolic growth rate during that passage; it need not have the same value at every point of the NHIM. An incoming initial condition has hyperbolic displacement proportional to $\sqrt I$. During the local passage that displacement grows approximately as
\begin{equation}
 \sqrt I\,e^{\Lambda t}.
\end{equation}
The trajectory leaves a chosen local region surrounding the NHIM when this quantity reaches a fixed scale $C$, so
\begin{equation}
 \sqrt I\,e^{\Lambda t_{\rm loc}}\sim C.
 \label{eq:Iloc}
\end{equation}
Taking logarithms gives
\begin{equation}
 t_{\rm loc}
 \simeq\frac{1}{\Lambda}\ln\!\left(\frac{C}{\sqrt I}\right),
 \qquad
 e^{\Lambda t_{\rm loc}}\propto I^{-1/2}.
 \label{eq:Idwell}
\end{equation}
The exponent $-1/2$ comes from the $\sqrt I$ initial hyperbolic displacement; it does not require identifying $\Lambda$ with the saddle-equilibrium eigenvalue at $E_s$.

Equation~\eqref{eq:Idwell} shows why smaller $I$ lengthens the local passage and increases the amplification accumulated during that passage. It is not, by itself, a formula for LB at fixed $\tau_*$. Equation~\eqref{eq:LB} combines both time directions, includes motion outside the local region, and averages over the split time. Both stretching histories start at the same dividing-surface point, so the observable includes local passage on both sides of that surface. Indeed, the exactly linear saddle has displacement-dependent passage times but spatially constant LB at fixed observation time.

\begin{figure}[ht]
\centering
\includegraphics[width=0.75\linewidth]{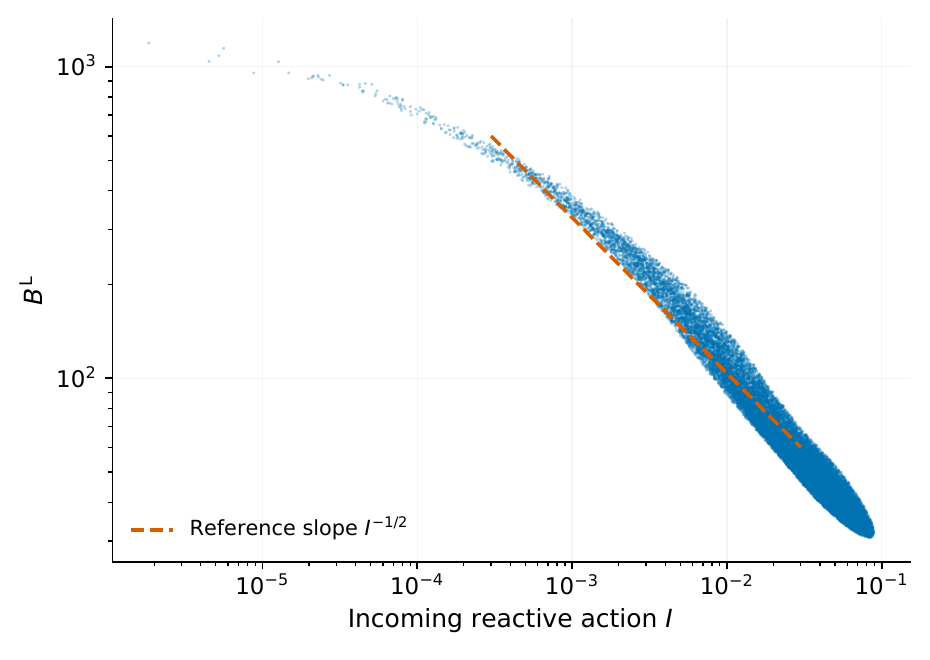}
\caption{Pre-exit LB at $\tau_*=0.0285$ ps versus the incoming reactive action $I$; both axes are logarithmic. Each point represents one incoming trajectory. The dashed line is a reference slope for the local passage-amplification scale in Eq.~\eqref{eq:Idwell}, not a fit to the ensemble. Moving toward smaller $I$ means choosing initial conditions closer to the NHIM; over most of the sampled range this produces larger pre-exit LB. The leveling at the smallest $I$ is consistent with the fixed observation interval (see text).}
\label{fig:LBaction}
\end{figure}

The molecular data support the association between smaller reactive action and larger pre-exit LB. A log--log fit of LB against $I$ over the full ensemble gives slope $-0.55$. Figure~\ref{fig:LBaction} displays that relation alongside the local passage-amplification scale. The fitted slope describes this ensemble over a fixed observation interval; Eq.~\eqref{eq:Idwell} does not derive a power law for the full LB integral.

For $I\lesssim10^{-4}$ the curve rolls over and approaches values near $1.2\times10^3$ at the smallest sampled actions. This is consistent with the finite observation interval. Once a trajectory remains locally hyperbolic for all of $\tau_*$, starting still closer to the NHIM cannot extend the time available to accumulate stretching. The leveling is analogous to the local flatness in Secs.~\ref{sec:localtheory} and~\ref{sec:benchmarks}, not an $I\to0$ divergence. It does not imply one common limiting LB value everywhere on the higher-energy NHIM, where the center motion and transverse rate may differ.

Smaller reactive action is associated with both larger pre-exit LB and longer gap times, and the local passage calculation provides a dynamical explanation for these tendencies. The evidence supports interpreting the LB--gap-time relation primarily in terms of transition-state-organized passage. It does not imply that the same early LB value ranks residence times associated with later trapping.

\subsection{Passage time is not the same as later trapping}

We now ask that separate question directly. Consider only trajectories that are still inside the molecular region at $0.5$ ps. Does the LB value measured much earlier, at $\tau_*=0.0285$ ps and before any trajectory had exited, tell us which of these already long-lived trajectories will leave sooner and which will leave later?

Both numerical settings identify the same 599 trajectories still inside at $0.5$ ps. We examine their subsequent exits over the interval ending at $5$ ps. This cutoff fixes the time range of the comparison: all eventual exits have been computed, including those beyond the cutoff. With the primary gap times, 487 of the 599 trajectories leave by $5$ ps and 112 leave later. The latter are excluded from this exit-time ranking. The refined calculation changes this partition but gives the same interpretation, as detailed in Appendix~\ref{app:numerics}.

For each of the 487 trajectories that exit within the chosen interval, we return to the \emph{original} pre-exit LB value measured at $\tau_*$. We rank these same trajectories once by that early LB value and once by their later exit time. The rank correlation is close to zero:
\begin{equation}
 \rho_S\!\left(B^{\rm L}(\tau_*),t_{\rm gap}\right)=-0.015,
 \qquad N=487.
 \label{eq:preexitlongrho}
\end{equation}
For these exits, pre-exit LB gives little useful indication of which trajectories leave earlier or later. A separate check using all 599 trajectories likewise finds that smaller versus larger early LB does not usefully distinguish which trajectories are still inside at $1$ ps; the definition and numerical values are given in Appendix~\ref{app:numerics}.

When stretching is accumulated through the full first $0.5$ ps, the forward FTLE does provide information about later survival: trajectories with weaker forward stretching are more likely to remain inside beyond $1$ ps, whereas LB over the same history remains much closer to chance. Appendix~\ref{app:numerics} gives the quantitative comparison. This association is consistent with residence in weakly stretching regular or sticky regions, but the comparison does not identify those regions or establish that mechanism.


\section{Discussion and Conclusions}
\label{sec:discussion}
\label{sec:conclusions}

An invariant manifold is defined by the dynamics. A scalar field computed from trajectory histories may indicate its influence, but the feature in the field still requires a dynamical explanation. For the linear saddle, LB is spatially constant even though the stable and unstable manifolds are present, whereas the $0<\alpha<1$ LD has an explicit nondifferentiable transverse dependence. FTLE and LB both derive from tangent amplification, but LB combines a family of backward and forward stretching factors over split times. Nonlinearity gives these quantities spatial variation without forcing an invariant manifold to be their singular set. Singular-value crossings provide a separate possible source of nondifferentiability, as noted in Sec.~\ref{sec:metric}.

The two limiting procedures explain why this distinction matters in practice. Under the two-sided local bounds of Sec.~\ref{sec:localtheory}, shrinking the initial neighborhood toward the saddle equilibrium makes the relative LB variation vanish even when the observation time follows the logarithmic local escape scale. Holding spatial resolution fixed and increasing observation time allows trajectories to accumulate different nonlinear histories, concentrating the spatial variation near manifold-organized transport. The quartic model displays both limits, and the coupled periodic-orbit model shows that local flattening persists when separability is lost.

There is also a long-time regime. Garc\'ia-Cuadrillero \emph{et al.} found that sufficiently long LB integrations in LiCN/LiNC are dominated by additional global structures rather than by the transition-state cross \cite{GarciaCuadrilleroEtAl2026}. A long observation window accumulates deformation from every region visited. Weakly stretching regular islands and strongly stretching chaotic regions can then develop a contrast larger than the localized transition-state signal. Increasing $\tau$ changes the question addressed by the diagnostic: the interval that makes a local bottleneck visible need not be the interval that distinguishes subsequent residence within the well.

The HCN/CNH calculations give this distinction a chemical interpretation. The slice centered at the equilibrium tests local regularity near the saddle at $E_s$; the reactive calculation uses the higher energy $E_s+0.2$ eV, where a three-dimensional NHIM and its dividing surfaces organize passage. On that surface, smaller incoming reactive action accompanies both longer gap times and larger pre-exit LB. The local passage calculation explains why reducing the displacement from the NHIM lengthens the passage and increases its amplification. The measured LB slope describes the combined forward and backward stretching history over the chosen finite interval, including motion outside the local region.

Among the selected long-lived trajectories, the original pre-exit LB gives little useful information about the ordering of later exits. The forward FTLE accumulated through $0.5$ ps is more informative about subsequent survival, showing that a longer stretching history can carry information absent from the original passage measurement. These are different dynamical questions, even though both involve residence in the same molecular region.

The network origin of LB helps explain the distinction. Betweenness identifies bridges between otherwise more internally connected regions \cite{GirvanNewman2002}. Passage through a transition-state bottleneck connects such regions; prolonged circulation within a well need not. Ser-Giacomi \emph{et al.} already distinguished source-to-destination passage from time spent in the intervening region \cite{SerGiacomiEtAl2021}. In the molecular example studied here, LB characterizes transport organized by independently known transition-state geometry. Its value lies in that description of passage, while its interpretation as a measure of later molecular residence requires a separate test.


\section*{Acknowledgments}

The recent LiCN/LiNC study of García-Cuadrillero \emph{et al.}~\cite{GarciaCuadrilleroEtAl2026} provided the immediate motivation for the questions examined here.

\section*{Author Declarations}

\subsection*{Conflict of Interest}
The author has no conflicts of interest to disclose.

\subsection*{Author Contributions}
Stephen Wiggins: Conceptualization; Methodology; Software; Validation; Formal analysis; Investigation; Data curation; Visualization; Writing (original draft); Writing (review and editing).

\section*{Data Availability}

The data, code, and computational documentation supporting this study are available on Zenodo at \url{https://doi.org/10.5281/zenodo.22874915} (Ref.~\onlinecite{WigginsRepro2026}). The canonical calculation records in this release support the present revision; its archived September 12 manuscript is retained as a historical snapshot. The package includes the normal form and its coordinate transformations, dividing-surface construction and sampling, gap-time and stretching calculations, and the numerical refinement records. The archived methods documentation gives the complete computational procedures; Appendix~\ref{app:numerics} reports the comparisons needed to assess the results.

\appendix

\section{Derivation of the local nonlinear estimate}
\label{app:proof}

This appendix supplies the steps behind Eq.~\eqref{eq:LBbound}. Let $M(t)=D\Phi^t(z_0)$ be the tangent map for Eq.~\eqref{eq:localvf}. It satisfies
\begin{equation}
 \dot M(t)=\left[A+DR(\Phi^t z_0)\right]M(t),\qquad M(0)=I.
 \label{eq:varapp}
\end{equation}
Variation of constants gives \cite{Walter1998}
\begin{equation}
 M(t)=e^{At}+\int_0^t e^{A(t-s)}DR(\Phi^s z_0)M(s)\,\dd s.
 \label{eq:duhamelapp}
\end{equation}
In the orthonormal saddle--center norm, $\norm{e^{At}}\le C_Ae^{\lambda t}$. While the trajectory remains in the local region, Eqs.~\eqref{eq:DRbound} and \eqref{eq:orbitbound} imply
\begin{equation}
 \norm{DR(\Phi^s z_0)}\le K C_0^\nu\delta^\nu e^{\nu\lambda s}.
\end{equation}
The constants must remain bounded as the observation time increases. For $0\le t\le\tau_\rho(\delta)$, the integrated correction satisfies
\begin{equation}
 \int_0^t\norm{DR(\Phi^s z_0)}\,\dd s
 \le \frac{K C_0^\nu\delta^\nu}{\nu\lambda}
       (e^{\nu\lambda t}-1)
 \le \frac{K\rho^\nu}{\nu\lambda}.
 \label{eq:integratedDRbound}
\end{equation}
The last bound is independent of $\delta$. Grönwall's inequality \cite{Walter1998} applied to $e^{-\lambda t}\norm{M(t)}$ therefore gives
\begin{equation}
 \norm{M(t)}\le C_M e^{\lambda t},\qquad
 C_M=C_A\exp\!\left(\frac{C_AK\rho^\nu}{\nu\lambda}\right).
 \label{eq:uniformMbound}
\end{equation}
Thus $C_M$ is uniform on the logarithmically growing interval for fixed $\rho$. Applying the same argument to the reversed vector field gives the backward bound, provided its trajectory obeys Eq.~\eqref{eq:orbitbound}. Substituting in Eq.~\eqref{eq:duhamelapp} gives
\begin{equation}
 \frac{\norm{M(t)-e^{At}}}{e^{\lambda t}}
 \le C\delta^\nu\frac{e^{\nu\lambda t}-1}{\nu\lambda}.
 \label{eq:Duhamelbound}
\end{equation}
The largest singular value is Lipschitz with respect to the operator norm,
\begin{equation}
 \abs{\sigma_{\max}(M)-\sigma_{\max}(N)}\le\norm{M-N}.
\end{equation}
Hence each stretching factor differs from its linear value by a relative correction of order $\delta^\nu(e^{\nu\lambda t}-1)$. In the product, the first-order terms involve durations $s$ and $\tau-s$. Up to a constant independent of $\delta$, their averaged size is bounded by
\begin{equation}
 \frac{\delta^\nu}{\tau}\int_0^\tau
 \left[(e^{\nu\lambda s}-1)
       +(e^{\nu\lambda(\tau-s)}-1)\right]\,\dd s
 =2\delta^\nu\left[
 \frac{e^{\nu\lambda\tau}-1}{\nu\lambda\tau}-1\right].
 \label{eq:splitaveragebound}
\end{equation}
The product of the two corrections is $\OO(\delta^{2\nu}e^{\nu\lambda\tau})$, because their durations add to $\tau$. These bounds give Eq.~\eqref{eq:LBbound}, including the $1/\tau$ reduction supplied by split-time averaging. No differentiability of the largest singular value is used in this estimate.

\section{Numerical checks and supporting details}
\label{app:numerics}

Two trajectories can agree during passage near a transition state and separate during subsequent chaotic motion. Numerical verification therefore has to follow the question being asked. The checks below distinguish the accuracy of the local event and stretching calculations from the sensitivity of the later gap times. They also specify which trajectories enter each statistical comparison.

\subsection{Numerical methods and refinement}
\label{app:reproduction}

The archive supplies the normal form, direct and inverse transformations, and fixed $25\,000$-point action-angle ensemble \cite{WigginsRepro2026}. The original random seed is unavailable, so reproduction uses the supplied samples and physical initial states; a separate seeded sampler supports new ensembles. The full algorithms, numerical controls, fit ranges, restart procedure, and validation records are documented there.

Both gap calculations propagate every incoming point independently from time zero, using the degree-ten exit condition in Sec.~\ref{sec:hcnmodel}. Primary DOP853 relative and absolute tolerances are $2\times10^{-12}$ and $2\times10^{-14}$, with maximum step $0.025\,t_{\rm u}$; refinement uses $2\times10^{-13}$, $2\times10^{-15}$, and $0.0125\,t_{\rm u}$. The largest recorded energy drifts are $4.36\times10^{-10}$ and $1.14\times10^{-11}$ eV, respectively.

Pre-exit stretching uses fourth-order Runge--Kutta integration with step $0.005\,t_{\rm u}$ and trapezoidal quadrature at spacing $0.01\,t_{\rm u}$. For the longer $0.5$-ps histories, the physical and tangent equations are integrated with DOP853 in both directions from the original incoming point, using the controls above and exact observation endpoints. Refined Simpson reductions supply the reported values. The largest relative LB change under integration refinement is $6.27\times10^{-5}$, with all diagnostic rank orderings unchanged; the largest nested-quadrature change is $1.56\times10^{-5}$ for LB and $9.76\times10^{-5}$ for the LD. Local-model averages use tensor-product Simpson weights. In Fig.~\ref{fig:hcndiff}, integration and quadrature refinement changes the LB slope differences by at most $0.104\%$ and the LD differences by $2.46\times10^{-11}$ relative.

\subsection{Sensitivity of the long gap-time tail}

Both calculations reach $25\,000$ observed exits, with no remaining censoring or solver failures:
\begin{center}
\begin{tabular}{l@{\qquad}r@{\qquad}r}
\toprule
quantity & primary & refined\\
\midrule
mean gap (ps) & 0.1603 & 0.1745\\
maximum gap (ps) & 158.703 & 418.899\\
channel-1 / channel-2 counts & 2533 / 22467 & 2525 / 22475\\
longest 100 gaps: share of total gap time & $49.21\%$ & $53.87\%$\\
\bottomrule
\end{tabular}
\end{center}
The mean changes by $8.89\%$, and 94 trajectories change exit channel. The longest-lived records contribute approximately half the total gap-time sum. Completion of this finite ensemble therefore does not establish convergence of individual late exits or of a population mean. By comparison, the largest difference between the empirical survival probabilities, the fractions with $t_{\rm gap}>c$ at cutoff $c$, is $0.00064$.

For sampling uncertainty, complete trajectory records are resampled with replacement $10\,000$ times, using seed 20260921 and paired identifiers across numerical settings \cite{EfronTibshirani1993}. The $2.5$ and $97.5$ percentiles give mean-gap intervals $[0.1373,0.1868]$ and $[0.1393,0.2215]$ ps. The interval for the refined-minus-primary mean is $[-0.0222,0.0606]$ ps. These intervals describe empirical sampling variability conditional on the computed outcomes; they do not include numerical bias or unobserved-tail uncertainty.

\subsection{Rank correlations}

The main text introduces Spearman's coefficient by its direct meaning: the same trajectories are sorted independently by two quantities, and the coefficient measures whether the resulting orderings are similar or reversed. Formally, Spearman's $\rho_S$ is the ordinary Pearson correlation applied to the two sets of ranks \cite{Spearman1904}. Thus $\rho_S=1$ corresponds to identical rank ordering, $\rho_S=-1$ to reversed ordering, and values near zero to no systematic monotonic relation between the rankings.

The first row of Table~\ref{tab:stretchcompare} gives the supporting pre-exit comparisons for all $25\,000$ trajectories. The primary LB--gap coefficient is $0.909$, with empirical interval $[0.907,0.911]$ from the resampling described above; refined gap labels change it by less than $2\times10^{-6}$. The reactive-action correlations are $\rho_S(I,t_{\rm gap})=-0.921$ and $\rho_S(I,B^{\rm L})=-0.976$. Additional conditioned LB--gap coefficients are $0.901$ for $t_{\rm gap}<0.05$ ps and $0.0058$ for $t_{\rm gap}\ge0.05$ ps. The channel-resolved coefficients are $0.075$ for channel 1 and $0.908$ for channel 2. Their populations and all six observation windows are recorded in the archive. The dashed guide in Fig.~\ref{fig:LBaction} is $k/\sqrt I$, where $k=10.38$ is the median of $B^{\rm L}\sqrt I$ over all 13\,595 points in the inclusive interval $3\times10^{-4}\le I\le3\times10^{-2}$, without additional filtering.

\subsection{Long-lived survivor check}

Both gap calculations select the same 599 trajectories with $t_{\rm gap}>0.5$ ps and the same 359 with $t_{\rm gap}>1$ ps. The primary/refined numbers surviving beyond $2$ ps are 217/218 and beyond $5$ ps are 112/100. All eventual exits have been computed. For the finite-time question in Sec.~\ref{sec:gaptimes}, we nevertheless impose an analysis cutoff at $5$ ps: the exit-time ranking uses the 487 primary trajectories with $0.5<t_{\rm gap}\le5$ ps. The other 112 are administratively censored for this comparison and are never assigned a fictitious $5$-ps gap. The refined calculation gives 499 observed exits and 100 administratively censored records within the same initial cohort.

Ranking the 487 observed exits by pre-exit LB and by their later exit times gives
\begin{equation}
 \rho_S\!\left(B^{\rm L}(\tau_*),t_{\rm gap}\right)=-0.0152.
\end{equation}
The empirical $95\%$ interval is $[-0.104,0.074]$. With refined gap labels the coefficient is $0.0105$ and the interval $[-0.077,0.096]$. Survivor intervals use $10\,000$ resamples with seed 20260922. Complete records from the 599-member cohort are resampled, and the observed-by-$5$-ps mask is reapplied within each resample. The two comparisons support the main-text conclusion that pre-exit LB supplies little useful monotonic ranking of these later exits.

A second question uses every member of the cohort: can a diagnostic distinguish trajectories that survive beyond $1$ ps from those that exit earlier? The area under the receiver-operating-characteristic curve (AUC) measures this rank separation, with $0.5$ representing chance and $1$ perfect separation \cite{HanleyMcNeil1982}. We fix weaker stretching as the positive survival score. The one-ps labels agree exactly between gap settings, so the corresponding AUCs and resampling intervals also agree.

\begin{table}[ht]
\caption{Supporting HCN/CNH comparisons. The first row uses all $25\,000$ pre-exit diagnostics and primary eventual gap times. The AUC rows use the same 599 trajectories surviving to $0.5$ ps and ask whether they survive beyond $1$ ps. Weaker stretching is scored as more long-lived; $0.5$ is chance. The last row uses the refined adaptive histories accumulated through exactly $0.5$ ps.}
\label{tab:stretchcompare}
\begin{ruledtabular}
\begin{tabular}{lccccc}
experiment & LB & $\lambda_f$ & $\lambda_b$ & $\lambda_f+\lambda_b$ & $M_{1/2}$\\
\hline
pre-exit $\rho_S$ with $t_{\rm gap}$ & 0.909 & 0.890 & 0.769 & 0.860 & 0.314\\
1-ps survival AUC from pre-exit values & 0.475 & 0.473 & -- & -- & --\\
1-ps survival AUC after 0.5 ps & 0.531 & 0.636 & -- & -- & --\\
\end{tabular}
\end{ruledtabular}
\end{table}

For pre-exit LB and forward FTLE, respectively, the AUC intervals are $[0.428,0.522]$ and $[0.426,0.521]$. When the diagnostics accumulate through the first $0.5$ ps, the intervals are $[0.482,0.578]$ for LB and $[0.591,0.681]$ for forward FTLE. Thus the longer forward-FTLE history is more informative about subsequent one-ps survival, whereas LB remains much closer to chance. The longer diagnostic includes deformation from a substantial part of the trajectory history and asks a different question from the original pre-exit measurement.

For the 487 observed exits, LB accumulated through $0.5$ ps has $\rho_S=-0.042$, with interval $[-0.129,0.049]$, and forward FTLE has $\rho_S=-0.259$, with interval $[-0.340,-0.172]$. Refined gap labels give $-0.064$ and $-0.279$, respectively. These comparisons concern diagnostic rankings within the stated cohort and cutoff. They do not establish statistical independence or exclude every nonmonotone relationship, and the observed-exit correlations are not correlations for the entire 599-member cohort.

\subsection{Coordinate change and the stretching norm}

Because LB is defined from singular values, its numerical value depends on the norm used to measure tangent vectors. If $y=\psi(z)$ is a smooth coordinate change, then
\begin{equation}
 D\widetilde\Phi^t(y)
 =D\psi(\Phi^t z)\,D\Phi^t(z)\,D\psi(z)^{-1}.
 \label{eq:coordchange}
\end{equation}
A canonical transformation is symplectic but need not be orthogonal. It therefore preserves the symplectic structure without necessarily preserving Euclidean lengths. If the metric is transformed together with the coordinates, the same geometric stretching is obtained in either coordinate system. If instead the Euclidean norm is imposed independently in each coordinate system, the numerical singular values generally change \cite{LekienRoss2010}.

\subsection{Metric comparison}

The full $25\,000$-trajectory pre-exit calculation was repeated after expressing the tangent maps in real linear normal-form coordinates and imposing the Euclidean norm there. For example, the median LB changes from $58.7$ in the physical mass-scaled metric to $31.1$ in the normal-form metric. The two LB rankings nevertheless have Spearman correlation $0.988$, and the LB--gap-time correlation changes only from $0.909$ to $0.898$. The numerical LB values therefore depend on the chosen norm, but the ranking of trajectories by LB changes very little in this calculation. This control uses the constant linear conjugation $S^{-1}MS$, with the same physical trajectories.

\end{document}